\documentclass[journal]{IEEEtran}

\usepackage{color}
\usepackage{amsmath}
\usepackage{amsfonts}
\usepackage{amsthm}
\usepackage{cite}
\usepackage{url}  
\usepackage[pdftex]{graphicx}
\usepackage{subcaption}
\usepackage{hyperref}
\usepackage{dsfont}
\hypersetup{
	colorlinks = true, 
	urlcolor  = cyan, 
	linkcolor = blue, 
	citecolor = red 
}
\usepackage{textcomp,mathcomp}
\usepackage{tabu}
\usepackage{booktabs}
\usepackage{makecell}
\usepackage{multirow}
\usepackage{threeparttable}

\graphicspath{{fig/}}

\newcommand{\relu}{\mathrm{ReLU}}

\begin{document}
	%
	\title{Coherent Hierarchical Probabilistic Forecasting of Electric Vehicle Charging Demand}

%
%
%

\author{Kedi~Zheng,~\IEEEmembership{Member,~IEEE,}
	Hanwei~Xu,
	Zeyang~Long,
	Yi~Wang,~\IEEEmembership{Member,~IEEE,}
	and~Qixin~Chen,~\IEEEmembership{Senior~Member,~IEEE}
	\thanks{This work was supported in part by the International (NSFC-NWO) Joint Research Project of National Natural Science Foundation of China under Grant 52161135201, in part by the Natural and Science Foundation of China under Grant 52307103, and in part by the Major Smart Grid Joint Project of the National Natural Science Foundation of China and State Grid under Grant U2066205. Paper no. 2023-KDSEM-1119. \textit{(Corresponding author: Qixin Chen.)}}
	\thanks{K. Zheng and Q. Chen are with the State Key Laboratory of Power Systems, Department of Electrical Engineering, Tsinghua University, Beijing 100084, China (e-mail: qxchen@tsinghua.edu.cn).}
	\thanks{H. Xu is with Beijing Deepseek Artificial Intelligence Fundamental Technology Research Co., Ltd., Beijing 100094, China.}
	\thanks{Z. Long is with School of Electrical and Electronic Engineering, Huazhong University of Science and Technology, Wuhan 430074, China.}
	\thanks{Y. Wang is with the Department of Electrical and Electronic Engineering, The University of Hong Kong, Hong Kong, China.}
	\thanks{Digital Object Identifier \href{https://doi.org/10.1109/TIA.2023.3344544}{10.1109/TIA.2023.3344544}}
}

\markboth{Manuscript for IEEE Transactions on Industry Applications, 2024}%
{Shell \MakeLowercase{\textit{et al.}}: Bare Demo of IEEEtran.cls for IEEE Journals}

\maketitle

\IEEEpubid{\begin{minipage}{\textwidth}\ \\[12pt] \centering
		© 2024 IEEE.  Personal use of this material is permitted.  Permission from IEEE must be obtained for all other uses, in any current or future media, including reprinting/republishing this material for advertising or promotional purposes, creating new collective works, for resale or redistribution to servers or lists, or reuse of any copyrighted component of this work in other works.
\end{minipage}}

\IEEEpubidadjcol

\begin{abstract}
	The growing penetration of electric vehicles (EVs) significantly changes typical load curves in smart grids. With the development of fast charging technology, the volatility of EV charging demand is increasing, which requires additional flexibility for real-time  power balance. 
	The forecasting of EV charging demand involves probabilistic modeling of high dimensional time series dynamics across diverse electric vehicle charging stations (EVCSs).
	This paper studies the forecasting problem of multiple EVCS in a hierarchical probabilistic manner. For each charging station, a deep learning model based on a partial input convex neural network (PICNN) is trained to predict the day-ahead charging demand's conditional distribution, preventing the common quantile crossing problem in traditional quantile regression models. Then, differentiable convex optimization layers (DCLs) are used to reconcile the scenarios sampled from the distributions to yield coherent scenarios that satisfy the hierarchical constraint. It learns a better weight matrix for adjusting the forecasting results of different targets in a machine-learning approach compared to traditional optimization-based hierarchical reconciling methods. Numerical experiments based on real-world EV charging data are conducted to demonstrate the efficacy of the proposed method.
\end{abstract}


\IEEEpeerreviewmaketitle

\section{Introduction}

\IEEEPARstart{T}{he number} of electric vehicles (EVs) is increasing rapidly worldwide in recent years. According to the Global EV Outlook, there were over 16.5 million EVs on the road in 2021, which was tripled in three years~\cite{IEA-2022}. The estimated electricity demand from EVs in 2030 would exceed 1400 TWh in the net zero emissions by 2050 scenario. While the EV electricity demand in China takes up 0.5\% of the country's final electricity demand in 2021, a conservative estimate of the number is expected to go beyond 3\% in 2030. Although there are discussions on whether fast charging or battery swapping should be the major pattern of future EV refueling solution~\cite{VALLERA2021112481}, fast charging is currently in a dominant position, and many obstacles need to be overcome for wider adoption of battery swapping mode. Thus, EV charging demand would undoubtedly be a significant component of the electricity load, with its special shape and volatility due to fast charging technologies.

EV charging station (EVCS) operators are important stakeholders in the era of EV, as they provide charging piles and services to general EV users in addition to their home chargers. An EVCS operator may own multiple geographically-distributed charging stations and operate them hierarchically in coordination with the grid~\cite{WU2022108005}. 
On many occasions, it may also be responsible for electricity purchase, ancillary service provision, and customer interaction~\cite{vagropoulos2013optimal,duan2021bidding,lyu2022electric}. 
Due to the stochastic nature of EV users, almost all the operating decisions of EVCS operators are made under uncertainty. 
Therefore, it is essential for EVCS operators to model the stochasticity and perform probabilistic forecasting of EV charging demand, generally in a hierarchical manner.

\IEEEpubidadjcol

\subsection{Related Work in Hierarchical Forecasting}

Hierarchical forecasting refers to the forecasting of multiple time series in hierarchy, i.e., some of the time series are the aggregation of others. Such hierarchical relationship of the time series would naturally yield a \textit{hierarchical constraint} (which will be further explained in Section~\ref{subsec:hierarchical-coherency}) for the values of the time series at any specific time point. However, individual forecasting of them cannot guarantee the \textit{hierarchical constraint}, leading to conflict of the results, reducing in accuracy, and further trouble in forecasting-guided operation and optimization.

At early stage, the \textit{hierarchical constraint} is enforced based on top-down or bottom-up forecasting~\cite{widiarta2009forecasting,Borges2013Evaluating}, i.e., distributing the high-level results to low-level or summing up the low-level results to high-level. However, such methods suffer from loss of information from individual series dynamics. Hyndman et al. proposed to use the reconciliation-based forecasting method to solve the hierarchical forecasting problem~\cite{HYNDMAN20112579,Wickramasuriya2019Optimal}. This method, a.k.a., coherent forecasting, solves an optimization problem of adjusting the base forecasting results of different levels so that the adjusted results satisfy the \textit{hierarchical constraint}. 

There are some emerging research related to hierarchical probabilistic forecasting. Taieb et al. \cite{taieb2017coherent,taieb2021hierarchical} proposed a bottom-up probabilistic forecast aggregation method based on copula theory. After aggregation, mean forecast combination and reconciliation are used to further ensure the coherency in the mean values. Hierarchical probabilistic load forecasting became the topic of the Global energy forecasting competition 2017 (GEFCom2017)~\cite{hong2019global}. Although several teams utilized the hierarchy information, few team discussed the problem of coherency at that time. Roach~\cite{roach2019reconciled} proposed an XGBoost-based reconciled forecasting model for GEFCom2017. However, the reconciliation mainly focused on the quantile forecasting results instead of the probablisitic distribution.

The forecasting results obtained from coherent forecasting has good statistical characteristics.
According to the empirical study of~\cite{PANAGIOTELIS2021343}, coherent forecasting usually improves the forecasting accuracy. It also ensures consistency of forecasted time series. Such advantages can further help the EVCS operator in consistent electricity purchase, efficient energy storage system operation, and effective customer interaction. 
However, there are still some open problems in coherent hierarchical forecasting, e.g., how to extend it to probabilistic forecasting~\cite{hong2020energy}, and how to implement it in a typical machine learning framework.

\subsection{Related Work in EV Demand Forecasting}

Although EV charging demand forecasting is a relatively new topic compared to traditional energy forecasting problems, e.g., load forecasting and renewable generation forecasting, there has been some noticeable work in recent years.

As for deterministic forecasting,
Arias et al.~\cite{ARIAS2016327} proposed a decision tree-based model to forecast the charging demand, utilizing historical traffic and weather data of the same region.
Saputra et al.~\cite{saputra2019energy} used a deep neural network (DNN) to forecast the energy demand of a certain area covering multiple charging stations. A federated learning approach is adopted to address the communication overhead and privacy issues.
Li et al.~\cite{li2022privacy} proposed a sophisticated federated learning framework for EVCS demand forecasting. The forecasting model is based on convolutional neural network (CNN), bi-directional long short-term memory (BiLSTM), and attention mechanism. Charging piles are divided into clusters, and inter-cluster and inner-cluster federated learning are used to train the base layer and the personalized layer of the model, respectively.
Dabbaghjamanesh et al.~\cite{Dabbaghjamanesh2021reinforcement} proposed a Q-learning based method for load forecasting of EVCS. Q-learning is a reinforcement learning technique that can generate more accurate forecasting based on conventional models of recurrent neural network (RNN) and artificial neural network (ANN).
Qiao et al.~\cite{Qiao2021data} forecasted the number of occupied charging piles at one certain EVCS using XGBoost.

As for probabilistic forecasting,
Huber et al.~\cite{HUBER2020114525} focused on performing quantile forecasts of EV parking duration and energy demand from its upcoming trip distance. A multi-layer perception-based quantile regression model and multivariate conditional kernel density estimators are applied. 
Buzna et al.~\cite{buzna2021ensemble} proposed an ensemble learning-based forecasting framework for hierarchical probabilistic EV load. Quantile regression-based methods such as linear quantile regression (LQR) and quantile regression
forests are used to generate forecasting results at low levels, and an $ \ell_1 $-penalized LQR model is used to ensemble the results and obtain the high-level results. 
Wu et al.~\cite{WU2022123475} used a parametric approach to model the arrival time and driven distances of EVs with normal and log-normal distributions. 
Hu et al.~\cite{hu2022self} combined the self-attention layers with the framework of machine theory of mind to perform quantile forecast. 
Li et al.~\cite{li2022probabilistic} used LSTM for point forecasting and Gaussian distribution for uncertainty modeling. A Markov decision process solved by a proximal policy optimization algorithm from reinforcement learning is used to forecast the variance of the Gaussian distribution. 

Several research gaps arise from the aforementioned analysis. The limited number of papers on probabilistic forecasting usually focus on the ultra-short-term forecast, e.g., 15-min ahead in~\cite{hu2022self} and 1-hour ahead in~\cite{li2022probabilistic}, which is exactly one time interval ahead depending on data granularity. Existing methods use parametric approaches (e.g., the Gaussian distribution) or the classical quantile regression to model the conditional distribution of EVCS demand. Parametric approaches assuming certain forms of potential distributions may not be able to capture the strong stochasticity in EV users' charging behaviors. In practice, classical quantile regression methods usually suffer from the annoying \textit{quantile crossing} problem. Moreover, few literature looked into the problem in a hierarchical way, which means the forecasting results do not satisfy the \textit{hierarchical constraint} in general.

\subsection{Contribution}
To address the research gaps in probabilistic modeling of EV charging demand and forecasting consistency among different multiple EVCSs, this paper proposes a novel deep learning-based forecasting framework. 
It adopts two types of neural network layers for convex learning. The first one is based on partial input convex neural network (PICNN)~\cite{pmlr-v70-amos17b,huang2021convex}, which learns a strictly convex function mapping the partial input to the output. The PICNN can be used to parameterize an invertible model for universal density approximation without \textit{quantile crossing}. It is used to construct a multi-horizon loss function and learn multi-variate distributions in forecasting problems~\cite{pmlr-v151-kan22a}. The second one is based on differentiable convex optimization layers (DCL)~\cite{NEURIPS2019_9ce3c52f}, which learns the mapping of parameters to optimal values of a certain type of convex optimization problems called disciplined parameterized programs. DCL can be adopted efficiently as a layer in back-propagation neural networks~\cite{pmlr-v70-amos17a}. We use PICNN to model the conditional distribution of EVCS demand in probabilistic forecasting, and DCL to further convert stochastic scenarios to coherent scenarios satisfying the \textit{hierarchical constraint}. 
The proposed method does not rely on any distributional assumptions of the target series.

The contribution of this paper is three fold:
\begin{enumerate}
	\item A novel framework based on deep learning for EV charging demand probabilistic forecasting is proposed. It solves practical issues of multi-variate stochasticity modeling and hierarchical coherency.
	\item PICNN is adopted to model the joint probabilistic distribution of multi-horizon EVCS demand as the gradient of a convex function w.r.t. quantile levels, which avoids potential problem of quantile crossing.
	\item DCL is introduced to further map individual forecasting results of different EVCS to coherent scenarios. The weight matrix for adjustment in hierarchical reconciliation can be learned in a deep learning manner.
\end{enumerate}
Note that discussion or innovation on the forecasting engines (e.g., ANN, RNN, LSTM, etc.) is beyond the scope of this paper. The forecasting engine used in the case study of this paper is based on the widely-adopted LSTM modules, which can be easily changed to other forecasting engines.

\subsection{Paper Structure}
The rest of this paper is structured as follows.
Section~\ref{sec:preliminary} presents some preliminary knowledge of the hierarchical and probabilistic forecasting problem.
Section~\ref{sec:framework} introduces the proposed EVCS demand forecasting framework. Section~\ref{sec:methodology} details the methodology of the adopted convex learning layers.
The case study is conducted in Section~\ref{sec:case}.
Finally, Section~\ref{sec:conclusion} draws the conclusion.

\section{Preliminaries}
\label{sec:preliminary}

\begin{figure}[t]
	\centering
	\includegraphics[width=0.95\linewidth]{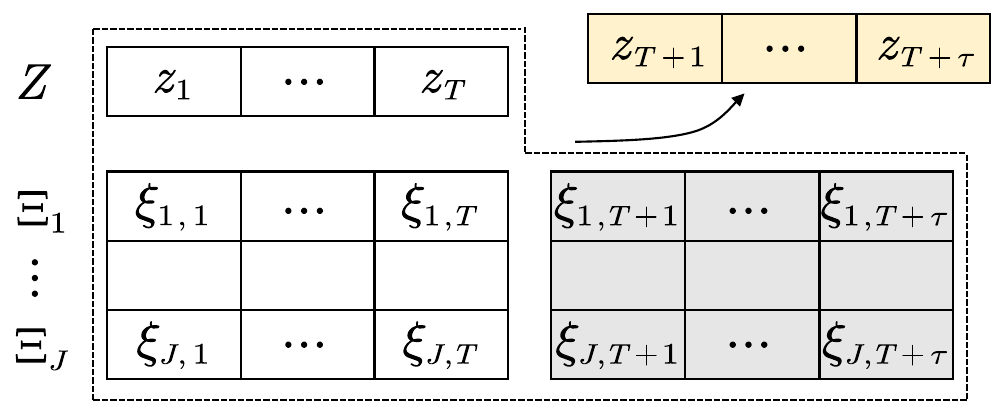}
	\caption{Relationship of observations in multi-horizon forecasting.}
	\label{fig:preliminary}
\end{figure}

\begin{figure*}[t]
	\centering
	\includegraphics[width=0.9\linewidth]{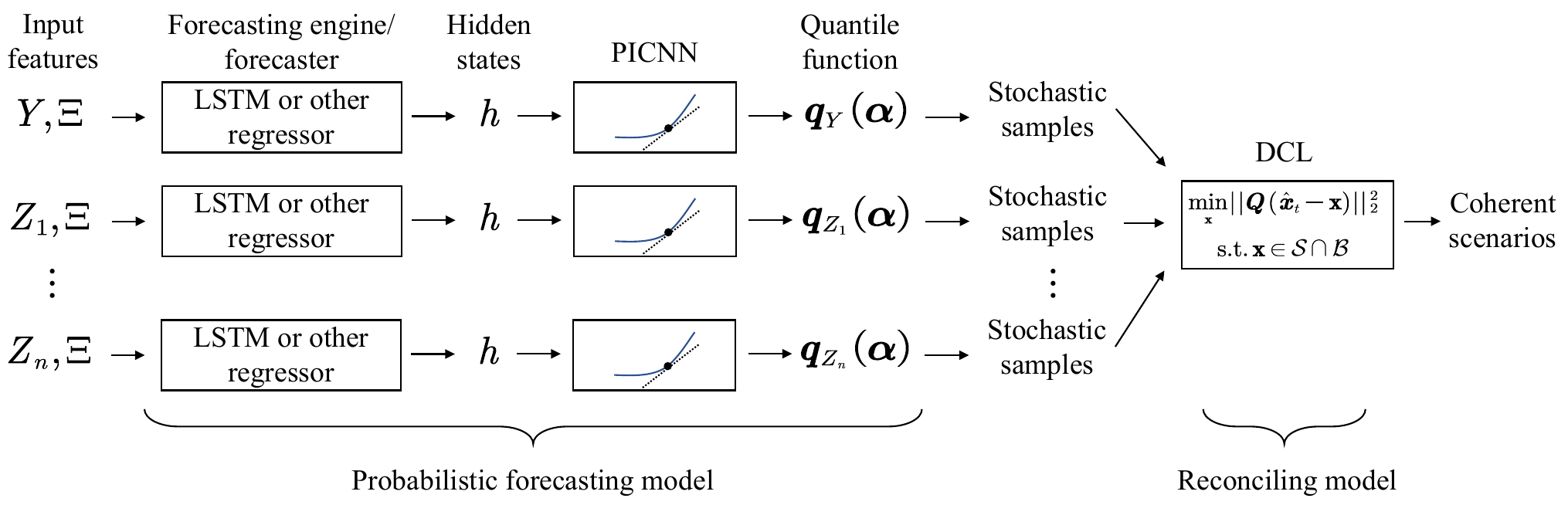}
	\caption{The proposed hierarchical probabilistic forecasting framework.}
	\label{fig:framework}
\end{figure*}

Let random variable $ Z \in \mathbb{R} $ denote the target series, and $ z_{t} $ denote the observation at different time $ t $. In a multi-horizon ahead forecasting problem, denote the current time as $ T $, and we want to forecast the future $ \tau $ observations at $ T+1,T+2,\cdots,T+\tau $. In addition to the target series, some covariates denoted as $ \Xi_j $ ($ j=1,\cdots, J $) that may have certain impact on the target series are also available as exogenous variables. The observations of $ \Xi_j $ are denoted as $ \xi_{j,t} $. Fig~\ref{fig:preliminary} shows the relationship of observations of different time series in a general multi-horizon forecasting problem. Depending on different settings, the future value of covariates may be accessible or not, which is marked as gray in Fig.~\ref{fig:preliminary}.
Although all observations before time $ T $ are available, a fixed length (a.k.a., context length) of the most recent observations would usually be used as input features for forecasting.

\subsection{Probabilistic Forecasting and Quantile Regression}
\label{subsec:quantile-crossing}

Probabilistic forecasting intend to predict the conditional distribution of $ z_{T+1}, \cdots, z_{T+\tau} $. Denote $ F_{Z}(z) $ as the cumulative distribution function (CDF) of $ Z $, and its quantile function is defined as:
\begin{equation}
	q_Z (\alpha) = F^{-1}_Z (\alpha) = \inf\{ z \in \mathbb{R} : \alpha \le F_Z(z) \}
\end{equation}
w.r.t. the quantile level $ \alpha \in (0,1) $. Since $ F_Z (\cdot) $ is monotonic increasing, $ q_Z (\alpha) $ should also be monotonic increasing for $ \alpha $.

Instead of learning a general function of $ q_Z (\alpha) $ for all time intervals, quantile regression uses forecasting engines to predict a conditional function $ q_{Z,t} (\alpha) $ from input features, for some given values of $ \alpha $. Awkward situation would arise if under certain input features, 
\begin{equation}
	q_{Z,t} (\alpha_1) > q_{Z,t} (\alpha_2), \ \text{even if}\ \alpha_1 < \alpha_2,
\end{equation}
which is named \textit{quantile crossing}. It indicates unreliable forecasting results and causes further trouble in scenario generation. Ideally, we desire to obtain:
\begin{equation} \label{equ:quantile-monotonicity}
	\left( q_{Z,t} (\alpha_1) - q_{Z,t} (\alpha_2) \right) (\alpha_1 - \alpha_2) \ge 0, \ \forall \alpha_{1,2} \in (0,1)
\end{equation}
regardless of input features. Since multi-horizon forecasting is conducted for $ \tau $ time intervals, (\ref{equ:quantile-monotonicity}) can be extended to a vector form:
\begin{equation}
	\label{equ:quantile-monotonicity-vector}
	\left( \boldsymbol{q}_{Z,t} (\boldsymbol{\alpha}_1) - \boldsymbol{q}_{Z,t} (\boldsymbol{\alpha}_2) \right) (\boldsymbol{\alpha}_1 - \boldsymbol{\alpha}_2) \ge 0, \ \forall \boldsymbol{\alpha}_{1,2} \in (0,1)^\tau
\end{equation}
Here $ \boldsymbol{q}_{Z,t} $ can be regarded as a mapping from $ (0,1)^\tau $ to $ \mathbb{R}^\tau $~\cite{pmlr-v151-kan22a}, in the sense of being a gradient of a convex function~\cite{Guillaume2016vector}.

\subsection{Hierarchical Coherency}
\label{subsec:hierarchical-coherency}

The notion of coherency in hierarchical forecasting has been well studied in a series of works by Hyndman et al~\cite{HYNDMAN20112579,Athanasopoulos2020}.
In the case of EV charging demand forecasting, the sum of the demand of multiple EVCSs should equal the total demand of the EVCS operator.
Let $ Z_i $ denote the time series of the demand of EVCS $ i $ at the bottom level ($ i = 1,\cdots, n $), and $ Y $ denote the time series of the total demand. Their observations should satisfy the following \textit{hierarchical constraint}:
\begin{equation} \label{equ:hierarchical-constraint}
	y_t =  \boldsymbol{S} \boldsymbol{z}_t 
\end{equation}
where $ \boldsymbol{S} = [1,1,\cdots,1] $. We use $ \hat{\cdot} $ to denote individual forecasting results (a.k.a., base forecasting), and let $ \boldsymbol{x}_t = [y_t, z_{1,t}, \cdots, z_{n,t}]^\top $. Since $ \hat{\boldsymbol{x}} $ may not satisfy~(\ref{equ:hierarchical-constraint}), coherent results reconciled from $ \hat{\boldsymbol{x}} $ are desired. Methods of forecasting reconciliation can leverage the results of base forecasters to obtain coherent and more accurate forecasting results. Traditional methods are based on least square linear regression, while~\cite{Erven2015game,pmlr-v139-rangapuram21a} presented a more general formulation based on quadratic programming:
\begin{equation}
	\label{equ:hierarchical-reconciling}
	\begin{aligned}
		\boldsymbol{x}_t^{\star} = & \arg \min_{\mathbf{x}} || \boldsymbol{Q}_{\text{r}} (\hat{\boldsymbol{x}}_t - \mathbf{x} ) ||_2^2   \\
		= & \arg \min_{\mathbf{x}} \left( \hat{\boldsymbol{x}}_t - \mathbf{x} \right)^\top \boldsymbol{Q} \left( \hat{\boldsymbol{x}}_t - \mathbf{x} \right) \\
		& \mathrm{s.t.}\ \mathbf{x} \in \mathcal{S} \cap \mathbb{R}_{+}^{n+1}
	\end{aligned}
\end{equation}
where $ \boldsymbol{Q} = \boldsymbol{Q}_{\text{r}}^\top \boldsymbol{Q}_{\text{r}} $ is the weight for adjustment, $ \mathcal{S} $ is the subspace defined by~(\ref{equ:hierarchical-constraint}), and $ \mathbb{R}_{+}^{n+1} $ is the nonnegative orthant.

$\boldsymbol{Q}_{\text{r}} $ and $ \boldsymbol{Q} $ are chosen as an identity matrix or according to the weighting factors of historical base forecasting error.

\section{Framework}
\label{sec:framework}

The proposed EV charging demand forecasting framework is shown in Fig.~\ref{fig:framework}. 

For each time series, i.e., the demand of each EVCS or the total demand of the EVCS operator, a probabilistic forecasting model based on LSTM modules and PICNN~\cite{pmlr-v70-amos17b,huang2021convex,pmlr-v151-kan22a} is constructed. 
The LSTM and PICNN are trained together using the energy score~\cite{gneiting2007Strictly} as loss function. As the forecasting engine, the LSTM modules would forward the input features into hidden states. Then, the PICNN would further forward the hidden states and quantile level $ \boldsymbol{\alpha} $ into a variable that is partially convex w.r.t. $ \boldsymbol{\alpha} $. Based on the differentiable and convex nature of the PICNN, the gradient of the variable w.r.t. $ \boldsymbol{\alpha} $ can be calculated as the quantile function, and stochastic samples of the forecasting horizon can further be drawn for each times series. 

The original stochastic samples are further forwarded to a reconciling model based on DCL. We also use the energy score to train the reconciling model for the parameter $ \boldsymbol{Q} $. The output of the reconciling model is the coherent scenarios.
The reason for applying reconciliation to stochastic samples instead of the original quantile functions or CDFs is that linear constraints of random variables would become complex relationships concerned with convolution operations among CDFs and PDFs of the variables. Such relationship is very difficult to be embedded into a typical machine learning framework.

\section{Methodology}
\label{sec:methodology}

\subsection{Feature Engineering}

In addition to the EVCS demand itself, covaraites of weather and calendar features are constructed. These features are commonly adopted in electricity demand forecasting tasks.

The weather features consist of the air temperature (\textcelsius), the dew point (\textcelsius), and hourly precipitation level (mm). 
The weather data used for feature construction are obtained from the Python API of Meteostat~\cite{Meteostat}, which provides access to open data from national weather services including the National Oceanic and Atmospheric Administration (NOAA).
Since the ACN data contains three EVCSs located in California, US, namely, Caltech, JPL (NASA's Jet Propulsion Laboratory), and Office001 (an office building located in the Silicon Valley area), their corresponding latitudes and longitudes are used to fetch the data.
The missing values in the weather data are filled using linear interpolation.

The calendar features consist of the Boolean indicator for US holiday, the Boolean indicator for weekday, sine and cosine features of the hour of the day and the hour of the year.

\subsection{LSTM Module}

\begin{figure}[t]
	\centering
	\includegraphics[width=0.8\linewidth]{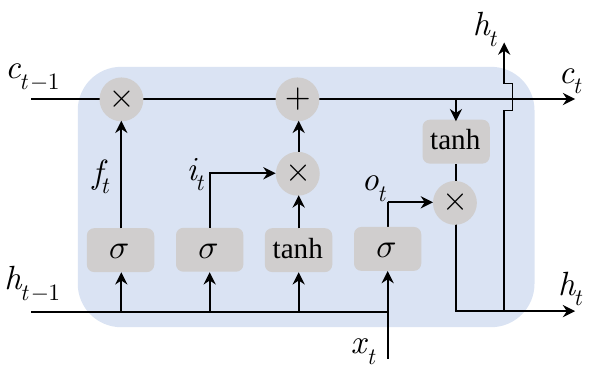}
	\caption{The structure of an LSTM module.}
	\label{fig:LSTM}
\end{figure}

We use the LSTM modules as the forecasting engine to extract hidden states for further probabilistic modeling. The structure of an LSTM module is shown in Fig.~\ref{fig:LSTM}. It has good capability in capturing temporal relationship between time series data. An LSTM module consists of a memory cell $ c_t $, a hidden state $ h_t $,  and three gates: the input gate $ i_t $ for extracting information from the input data for the memory cell, the forget gate $ f_t $  for discard historical information from past memory cell, and the output gate $ o_t $ for extracting information to further compute the hidden state $ h_t $. The three gates are typical neurons with weight and bias parameters as well as a sigmoid activation function $ \sigma $. The LSTM module calculates the following forward function:
\begin{equation}
	\begin{aligned}
		i_{t} & = \sigma \left( W^{(i,x)} x_t + W^{(i,h)}, h_{t-1} + b_i \right) \\
		f_t  & = \sigma \left( W^{(f,x)} x_t + W^{(f,h)} h_{t-1} + b_f \right)  \\
		o_t & = \sigma \left( W^{(o,x)} x_t + W^{(o,h)} h_{t-1} + b_o \right) \\
		c_t  & = i_t \otimes  \tanh \left( W^{(c,x)} x_t + W^{(c,h)} h_{t-1} + b_c \right) + f_t \otimes c_{t-1} \\ 
		h_t & = o_t \otimes \tanh  \left( c_t \right)
	\end{aligned}
\end{equation}
where $ W^{(\cdot)} $ and $ b^{(\cdot)} $ are weight and bias of the gate neurons. 

The LSTM modules can be replaced by any other machine learning regressors, e.g., multi-layer perceptions.

\subsection{Probabilistic Modeling with PICNN}

\begin{figure}[t]
	\centering
	\includegraphics[width=0.9\linewidth]{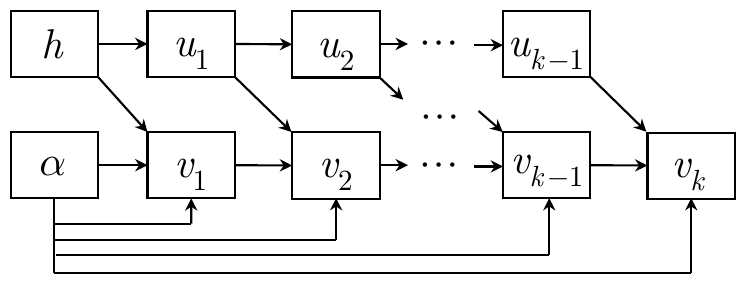}
	\caption{The structure of a multi-layer PICNN.}
	\label{fig:PICNN}
\end{figure}

The structure of a PICNN is shown in Fig.~\ref{fig:PICNN}, which fits a partial convex function on $ (\alpha, h) $. The forward function of each layer is:
\begin{equation}
	\begin{aligned}
		u_{i+1}  = & g^{(u)}_i \left( W^{(u,u)}_i u_i + b^{(u,u)}_i \right) \\
		v_{i+1}  = &  g^{(v)}_i \left( W_i^{(v)} \left( v_i \otimes \relu (W_i^{(v,u)} u_i + b_i^{(v,u)} ) \right) \right. \\
		&  \left. + W_i^{(\alpha)} \left( \alpha \otimes \relu ( W_i^{(\alpha,u)} u_i + b_i^{(\alpha,u)} ) \right)    \right. \\
		& \left.  +  W_i^{(u)} u_i + b_i^{(v)}  \right)
	\end{aligned}
\end{equation}
where $ i = 0, \cdots, k-1$ is the index for layers, $ g_{i}^{(\cdot)} $ is some activation function for layer $ i $, and $ u_0 = h, v_0 = \alpha $.
$ \relu (\cdot) = \max\{0, \cdot \} $ is the ReLU function.
The output of the PICNN is $ f(\alpha,h) = v_k $, which is convex on $ \alpha $ but not necessarily convex on $ h $, provided that $ W_i^{(v)} $ and $ W_{i}^{(\alpha)} $ are non-negative and $ g_i^{(v)} $ is convex non-decreasing. A brief proof of the \textit{partial convexity} is given here:
\begin{proof}
	The composition of a convex function and a convex non-deceasing function is also convex. 
	
	In the base case, $ v_0 $ is convex non-deceasing on $ \alpha $. Suppose $ v_i $ is convex non-deceasing on $ \alpha $. It is clear that the first term $ W_i^{(v)} \left( v_i \otimes  \relu (W_i^{(v,u)} u_i + b_i^{(v,u)} ) \right)  $ is convex non-deceasing on $ \alpha $, since $ \relu (\cdot) $ is non-negative. The second term $ W_i^{(\alpha)} \left( \alpha \otimes \relu ( W_i^{(\alpha,u)} u_i + b_i^{(\alpha,u)} ) \right)  $ positive linear on $ \alpha $. Since $ g_i^{(v)} $ is convex non-decreasing, $ v_{i+1} $ is convex non-decreasing on $ \alpha $.
	
	Using Mathematical Induction, we can conclude that $ v_{k+1} $ is convex non-decreasing on $ \alpha $.
\end{proof}

Various activation functions can be used in PICNN, including ReLU and Gaussian softplus. While no activation function dominates in all cases, usually ReLU performs well in fitting piece-wise functions, whereas softplus-like functions perform well in capturing nonlinearity. Therefore, it would be better to combine ReLU with softplus-like functions (see Appendix~\ref{sec:appendix} for more discussion). 

Once the convex mapping is established by PICNN, the quantile function can be further obtained through gradient:
\begin{equation}
	\begin{aligned}
		q \left(\alpha | h \right) & = \nabla_\alpha f \left(\alpha , h \right) = \frac{\partial v_k}{\partial \alpha} \\
		\boldsymbol{q} \left( \boldsymbol{\alpha} | \boldsymbol{h} \right) & = \nabla_{\boldsymbol{\alpha}} \boldsymbol{e}^\top \boldsymbol{f}\left( \boldsymbol{\alpha}, \boldsymbol{h} \right) = \nabla_{\boldsymbol{\alpha}} \boldsymbol{e}^\top \boldsymbol{v}_k
	\end{aligned}
\end{equation}
for the scalar and vector form, respectively. As stated in the end of Section~\ref{subsec:quantile-crossing}, since $\boldsymbol{q} \left( \boldsymbol{\alpha} | \boldsymbol{h} \right)$ is the gradient of a convex function, it should satisfy (\ref{equ:quantile-monotonicity-vector}) and avoid potential \textit{quantile crossing} problem. 

The energy score (ES), which is an extension of the continuous ranked probability score (CRPS) in the multivariate case~\cite{gneiting2007Strictly} and has been widely used for scenario evaluation in energy forecasting problems~\cite{pinson2012Evaluating,zheng2022data}, is adopted as the loss function for the training of the LSTM and PICNN modules. For some observation $ \boldsymbol{x} $, let $ \Omega $ denote the sample (scenario) set, and the energy score is calculated as:
\begin{equation} \label{equ:energy-score}
	L_{\text{ES}} = \mathbb{E}_{\boldsymbol{w}_1, \boldsymbol{w}_2 \in \Omega }  \left( -\frac{1}{2}\left|\left|   \boldsymbol{w}_1 - \boldsymbol{w}_2 \right|  \right|_2^\beta + \left|  \left|  \boldsymbol{w}_1 - \boldsymbol{x} \right| \right|_2^\beta  \right)
\end{equation}
where the parameter $ \beta \in \left( 0,2 \right) $.
$ \Omega $ is obtained by sampling a finite number of $ \boldsymbol{\alpha} $ from the uniform distribution:
\begin{equation}
	\Omega = \left\{ \boldsymbol{q}  \left(  \boldsymbol{\alpha} |  \boldsymbol{h} \right)   \  \big|  \  \boldsymbol{\alpha} \sim U\left(0,1\right)^ \tau \right\}
\end{equation}
Once the loss function is defined, the parameters of the probabilistic forecasting neural network can be optimized using stochastic gradient descent or its variants.

\subsection{Principle of DCL}

This paper adopts the DCL proposed in~\cite{NEURIPS2019_9ce3c52f} that aims to solve the gradient of the optimal solution of a special type of convex optimization problem called disciplined parametrized programming (DPP) w.r.t. the input parameters. DPP guarantees that the produced program can be reduced to affine-solver-affine (ASA) format, where affine stands for affine mapping and solver stands for a general conic solver. 

By introducing auxiliary variables, the quadratic programming problem of (\ref{equ:hierarchical-reconciling}) can be canonicalized into a a convex cone program of the form:
\begin{equation}
	\label{equ:cone}
	\begin{aligned}
		& \min_{\mathbf{x}}\ \boldsymbol{c}^\top \mathbf{x} \\
		& \mathrm{s.t.} \ \boldsymbol{b} - \boldsymbol{A}\mathbf{x} \in \mathcal{K} 
	\end{aligned}
\end{equation}
where $ \mathbf{x} $ is the variable, $ \mathcal{K} $ is a nonempty, closed, convex cone, and $\boldsymbol{A}$, $\boldsymbol{b}$, $\boldsymbol{c}$ are the input data of the problem. In (\ref{equ:hierarchical-reconciling}), omit the subscript $ t $, and let auxiliary variable $\boldsymbol{\xi} = \hat{\boldsymbol{x}} $ and $ \eta \ge || \boldsymbol{Q}_{\text{r}} (\boldsymbol{\xi} - \mathbf{x} ) ||_2 $. Thus, (\ref{equ:hierarchical-reconciling}) is equivalent to:
\begin{equation}
	\begin{aligned}
		\min_{\eta,\boldsymbol{\xi},\mathbf{x} } \ &   \eta \\
		\mathrm{s.t.}\ &  \left(\eta, \boldsymbol{Q}_{\text{r}} (\boldsymbol{\xi} - \mathbf{x}) \right) \in \mathcal{Q}_{n+2} \\
		& \boldsymbol{\xi} = \hat{\boldsymbol{x}} \\
		& \mathbf{x} \in \mathcal{S} \cap \mathbb{R}_{+}^{n+1}
	\end{aligned}
\end{equation}
Corresponding to (\ref{equ:cone}), here $ \left( \eta, \boldsymbol{\xi},\mathbf{x} \right) $ is the variable, $ \mathcal{Q}_{n+2} $ is the (n+2)-dimensional second-order cone, and the input data of the problem is:
\begin{equation}
	\begin{gathered}
		\boldsymbol{A} = \begin{bmatrix}
			-1  & 0 & 0 \\
			0  & -\boldsymbol{Q}_{\text{r}} & \boldsymbol{Q}_{\text{r}}  \\
			\cmidrule(lr){1-3}
			0	& -\boldsymbol{I} & 0 \\ 
			\cmidrule(lr){1-3}
			0	& 0	& -\boldsymbol{I}
		\end{bmatrix}, \quad \boldsymbol{b} = \begin{bmatrix}
			0\\
			0\\
			-\hat{\boldsymbol{x}} \\
			0
		\end{bmatrix}, \quad \boldsymbol{c} = \begin{bmatrix}
			1 \\
			0 \\
			0 \\
			0
		\end{bmatrix} \\
		\mathcal{K} = \mathcal{Q}_{n+1} \times \{0\} \times ( \mathcal{S} \cap \mathbb{R}_{+}^{n+1} )
	\end{gathered}
\end{equation}
where horizonal lines correspond to the cones in $ \mathcal{K} $. Thus, the input data $ (\boldsymbol{A}, \boldsymbol{b}, \boldsymbol{c}) $ are affine to $ (\boldsymbol{Q}_{\text{r}}, \hat{\boldsymbol{x}}) $.

The derivative of a conic solver for~(\ref{equ:cone}) can be solved as described in Section~4.3 of~\cite{NEURIPS2019_9ce3c52f}, and the solution retrieval from the conic solver to the optimal value $ \boldsymbol{x}^{\star} $ is also affine to $ (\boldsymbol{Q}_{\text{r}}, \hat{\boldsymbol{x}}) $. Therefore, the optimal solution of (\ref{equ:hierarchical-reconciling}) is differentiable w.r.t. $ \boldsymbol{Q}_{\text{r}} $, and the gradient for a typical differentiable loss function evaluated on $ \boldsymbol{x}^{\star} $ w.r.t. $ \boldsymbol{Q}_{\text{r}} $ can be calculated based on the chain rule. 

DCL has been implemented in a Python package called Cvxpylayers\footnote{\url{https://github.com/cvxgrp/cvxpylayers}}, which is compatible with the widely-used deep learning framework Pytorch, making it easy for embedding the gradient calculation process in end-to-end learning. 

\subsection{Trainable Hierarchical Reconciling with DCL}
\label{subsec:DCL-for-reconciling}

The general hierarchical reconciling problem defined in~(\ref{equ:hierarchical-reconciling}) has two parameters $ \hat{\boldsymbol{x}} $ and $ \boldsymbol{Q} $ (or $ \boldsymbol{Q}_{\text{r}} $). While $ \hat{\boldsymbol{x}} $ is the input from stochastic samples provided by the preceding probabilistic forecasting model, $ \boldsymbol{Q} $ is usually fixed in classical hierarchical reconciling methods as stated in Section~\ref{subsec:hierarchical-coherency}. 

It is straightforward to consider $ \boldsymbol{Q}_{\text{r}} $ as a trainable parameter in a general machine learning model. Given that $ \boldsymbol{Q}_{\text{r}} $ is the weight for adjustment and directly influences the reconciling outcomes, learning $ \boldsymbol{Q}_{\text{r}} $ from historical time series data and base forecasting results would be beneficial for obtaining more accurate forecasting results, and can be regarded as an extension of the traditional weighting factors estimation methods. 

\begin{figure}[!t]
	\centering
	\begin{subfigure}[t]{1.0\linewidth}
		\centering\includegraphics[width=0.8\linewidth]{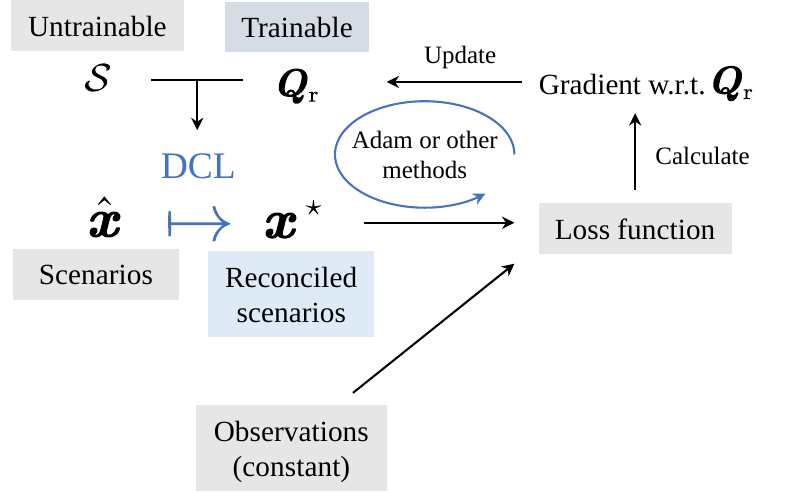}
		\caption{The general gradient descent-based training procedure.}
		\label{subfig:DCL-training}
	\end{subfigure}
	\\
	\begin{subfigure}[t]{1.0\linewidth}
		\centering\includegraphics[width=0.9\linewidth]{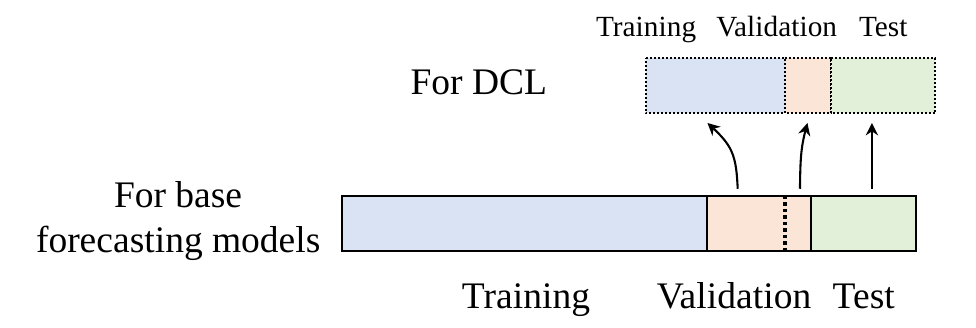}
		\caption{The chronological data partitioning.}
		\label{subfig:CRs}
	\end{subfigure}
	\caption{Training and data partitioning of hierarchical reconciling.}
	\label{fig:training-process}
\end{figure}

Fig.~\ref{fig:training-process} shows the training and data partitioning of hierarchical reconciling with trainable parameters. The DCL maps the original scenarios obtained from base forecasting models to the reconciled scenarios which is further used to compute a scenario-based loss (e.g., the energy score). The gradient of the loss w.r.t. $ \boldsymbol{Q}_{\text{r}} $ is then calculated, and $ \boldsymbol{Q}_{\text{r}} $ can be trained in typical stochastic gradient-based approaches. Only a part of DCL's parameters are trainable, as $ \mathcal{S} $ defining the hierarchical constraints is fixed. 
The dataset splitting for hierarchical reconciling has one more step in addition to the training, validation, and test set splitting for base forecasting models, i.e., the base validation set is split chronologically into the training and validation set for DCL. The test set remains the same for both DCL and base forecasting models.

\section{Case Study}
\label{sec:case}

To demonstrate the performance of the proposed hierarchical probabilistic forecasting method for EV charging demand, a case study is conducted based on the the ACN dataset~\cite{lee2019acn}. All the numerical experiments are conducted on a 36-core 3.00 GHz Ubuntu server with a NVIDIA GeForce RTX 3080 GPU. Python 3.8, Pytorch 1.13, Cvxpylayers 0.1.5, and CUDA 11.7 are used for the constructing and training of deep learning models. The Adam optimizer is used, the learning rate is set as 0.001, the batch size for training is 64, and the maximum number of epochs is 200. The length of historical series for forecasting, i.e., $ T $ in Fig.~\ref{fig:preliminary}, is selected as $ 7\times 24 $. 

\subsection{Data}

\begin{figure}[h]
	\centering
	\includegraphics[width=1.0\linewidth]{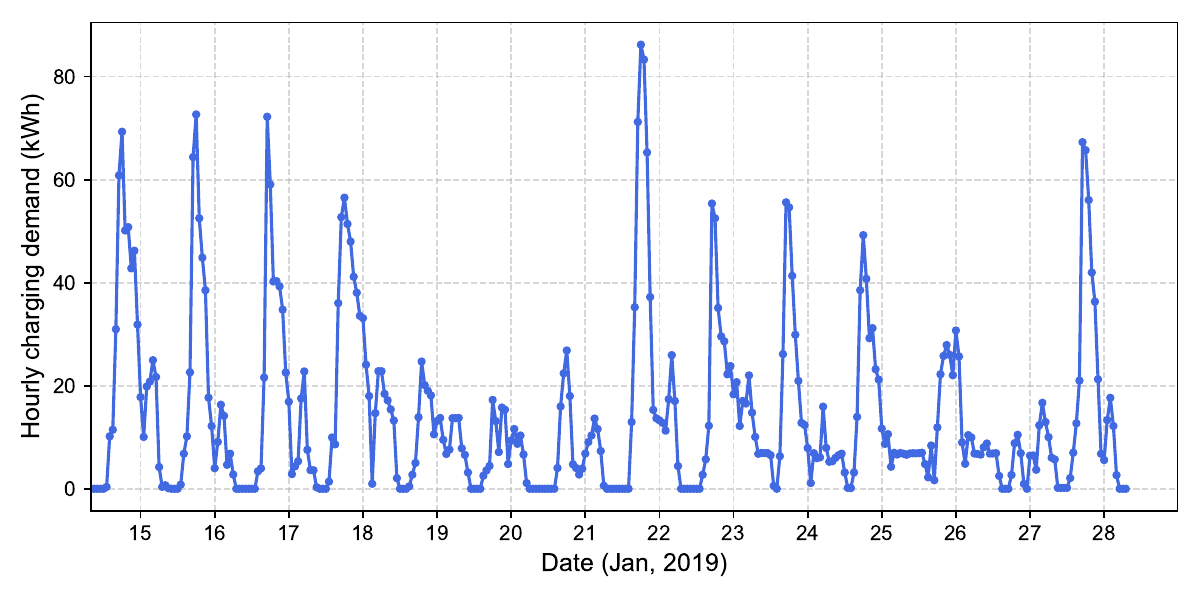}
	\caption{Two-week charging demand of the Caltech EVCS.}
	\label{fig:EVCS-demand}
\end{figure}

The ACN dataset contains three EVCSs named \texttt{Caltech}, \texttt{JPL}, and \texttt{Office001} in California. It records the amount of delivered energy (unit: kWh) for each anonymous charging session, as well as the start/end time and the charging pile ID. We aggregate data from all sessions and obtained the hourly charging demand for each EVCS. An example of the demand curve for the Caltech EVCS is shown in Fig.~\ref{fig:EVCS-demand}. As the three EVCSs are located in school and work areas, the data from January 15, 2019 to March 15, 2020 are used to construct the dataset for avoiding the impact of COVID-19 lockdown. The charging demand of each individual EVCS and the total demand of the three EVCSs (denoted as \texttt{Total}) are forecasted.

For the probabilistic forecasting model (the base model), the first 12-month data are used as the training set, the subsequent one-month data are designated as the validation set, and the final one-month data are used as the test set. 
For the reconciling model (DCL), the aforementioned validation set is further split into 80\% and 20\% for training and validation, respectively.

\subsection{Comparisons and Evaluation Metrics}

Three probabilistic forecasting methods are adopted as comparisons: MLP (Multi-Layer Perception), DeepAR~\cite{DeepAR}, and DeepVAR~\cite{DeepVAR}. 
MLP is a common method for point forecasting problems. It uses a fully-connected neural network to forecast the target series. 
DeepAR is a state-of-the-art method for multi-horizon probabilistic forecasting. Based on RNN modules, it predicts the stochastic scenarios of the target series in a auto-regressive manner, i.e., the method leverages both the historical observations of the series and previously predicted values to generate forecasts for subsequent time steps. 
DeepVAR is an extension of DeepAR for forecasting multivariate time series with a single model. 
The structure of PICNN in the proposed framework is based on the code of \cite{huang2021convex} available on Github\footnote{\url{https://github.com/CW-Huang/CP-Flow}}. 

For each model, 24-hour ahead forecasting is performed for each time interval in the validation and test set. 

A good point forecast usually indicates a good probabilistic forecast, and there is a positive correlation between point forecasting metrics and probabilistic forecasting metrics~\cite{wang2018conditional}. 
Thus, although the final output of the proposed method is probabilistic scenarios, both point forecasting metrics and probabilistic forecasting metrics are adopted to evaluate the results comprehensively. The three point forecasting metrics are:
\begin{enumerate}
	\item MAE (Mean Absolute Error): the average of the absolute differences between the predicted values and the actual observed ones. The formula for MAE is:
	\begin{equation}
		\text{MAE} = \frac{1}{n} \sum_{t} |x_t - \hat{x}_t|
	\end{equation}
	where $ n $ is the number of time intervals for evaluation.
	\item RMSE (Root Mean Squared Error): the square root of the mean squared error. The formula for RMSE is:
	\begin{equation}
		\text{RMSE} = \sqrt{\frac{1}{n} \sum_{t} (x_t - \hat{x}_t)^2 }
	\end{equation}
	\item MASE (Mean Absolute Scaled Error): the mean of the absolute differences between the predicted values and the actual observed values, scaled by the mean absolute error of a naïve forecast (usually a previous observation). The formula for MASE is:
	\begin{equation}
		\text{MASE}(t_0) = \frac{\sum_{t} | x_t - \hat{x}_t |}{n}  / \frac{\sum_{t} | x_t - x_{t-t_0} |}{n}
	\end{equation}
	The value of $ t_0 $ is selected as 24 and 168.
\end{enumerate}

The two probabilistic forecasting metrics are:
\begin{enumerate}
	\item Quantile Loss (QL): also known as pinball loss, which is used to assess the accuracy of quantile forecasts. The formula for QL is:
	\begin{equation}
		\begin{aligned}
			\text{QL}(\alpha) = & \frac{1}{n} \Big(\sum_{t} (1-\alpha) \cdot (\hat{x}_t - x_t) \cdot \mathds{1}(x_t < \hat{x}_t) \\ 
			& + \alpha \cdot (x_t - \hat{x}_t ) \cdot \mathds{1}(x_t \ge \hat{x}_t) \Big)
		\end{aligned}
	\end{equation}
	where $\mathds{1} (\cdot) $ is the boolean function for a certain condition, and $\alpha$ is the quantile level.
	\item Winkler Score (WS): used for assessing the accuracy of prediction intervals, considering both the width of the prediction interval and the coverage. The formula for WS is:
	\begin{equation}
		\begin{gathered}
			\text{WS}_{\alpha,t} = \begin{cases}
				L_t + \frac{2}{1-\alpha} ( \hat{x}_t^{\text{lower}} - x_t ) & \text{if } x_t < \hat{x}_t^{\text{lower}} \\
				L_t & \text{if } x_t \in [\hat{x}_t^{\text{lower}}, \hat{x}_t^{\text{upper}}] \\
				L_t + \frac{2}{1-\alpha} ( x_t - \hat{x}_t^{\text{upper}}) & \text{if } x_t > \hat{x}_t^{\text{upper}}
			\end{cases} \\
			\text{WS}(\alpha) = \frac{1}{n} \sum_{t} \text{WS}_{\alpha,t}
		\end{gathered}
	\end{equation}
	where $[\hat{x}_t^{\text{lower}}, \hat{x}_t^{\text{upper}}]$ is the predicted interval, $L_t = \hat{x}_t^{\text{upper}} - \hat{x}_t^{\text{lower}}$ is the length, and $ \alpha $ is the confidence level for the interval.
\end{enumerate}
Also, the energy score defined in (\ref{equ:energy-score}) is used to evaluate the stochastic scenarios before and after hierarchical reconciling. 

For MLP, the number of layers is selected between 2 to 4, and the size of hidden neurons are selected between 100 to 250, according to the loss in the validation set. To obtain probabilistic forecasting results, the MLP is trained in a quantile regression way using a summed QL function for all $ \alpha \in \left(  0.05,0.1,0.2,0.3,0.4,0.5,0.6,0.7,0.8,0.9,0.95 \right)$. For DeepAR and DeepVAR, the number of LSTM layers is 2, and the size of hidden neurons is 100. For the proposed method, the number of LSTM layers is also 2, and the size of hidden neurons is also 100. The number of layers for PICNN is 2, and the hidden size is 40. The activation function for each layer of PICNN is set as ReLU and Gaussian softplus, respectively.

\begin{table}[!t]
	\renewcommand{\arraystretch}{1.3}
	\caption{Point Forecasting Evaluation}
	\label{table:point-forecasing}
	\centering
	\begin{tabu}{cccccc}
		\toprule
		EVCS & Method & MAE & RMSE & MASE(24) & MASE(168) \\ \midrule
		\multirow{4}{*}{Caltech} & MLP & 4.9952 &  9.1214 &   0.6883 &    0.7869 \\
		& DeepAR & 3.6892 &  6.9941 &   0.5083 &    0.5811 \\
		& DeepVAR & 3.7841 &  7.0686 &   0.5214 &    0.5961 \\
		& Proposed &  \textbf{0.7981} &  \textbf{1.7804} &   \textbf{0.1100} &    \textbf{0.1257} \\ \hline
		\multirow{4}{*}{JPL}	 & MLP & 8.9547 & 16.9270 &   0.5148 &    0.6969 \\
		& DeepAR & 9.6658 & 17.7450 &   0.5557 &    0.7523 \\
		& DeepVAR & 5.7988 &  9.7675 &   0.3334 &    0.4513 \\
		& Proposed & \textbf{2.3712} &  \textbf{3.6923} &   \textbf{0.1363} &   \textbf{0.1845} \\  \hline
		\multirow{4}{*}{Office001}	 & MLP & 1.3682 &  2.8981 &   0.7220 &    1.1064 \\
		& DeepAR & 1.5436 &  2.7632 &   0.8145 &    1.2482 \\
		& DeepVAR & 0.9279 &  1.9140 &   0.4896 &    0.7503 \\
		& Proposed & \textbf{0.1825} &  \textbf{0.4330} &   \textbf{0.0963} &    \textbf{0.1476} \\ \hline
		\multirow{4}{*}{Total}   & MLP & 13.4911 & 25.8588 &   0.5886 &    0.8057 \\
		& DeepAR & 9.6734 & 16.2320 &   0.4221 &    0.5777 \\
		& DeepVAR & 7.4028 & 12.6054 &   0.3230 &    0.4421 \\
		& Proposed & \textbf{2.1546} &  \textbf{3.3495} &   \textbf{0.0940} &  \textbf{0.1287} \\						 
		\bottomrule
	\end{tabu}
\end{table}

\begin{figure}[t]
	\centering
	\begin{subfigure}[t]{1.0\linewidth}
		\centering\includegraphics[width=1.0\linewidth]{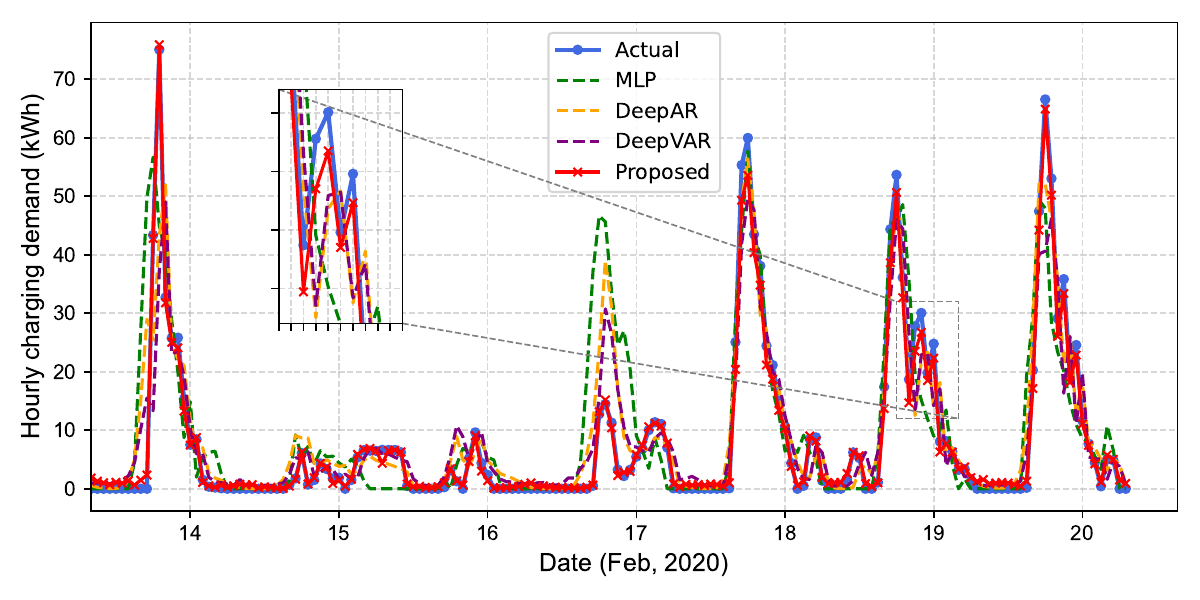}
		\caption{Caltech.}
		\label{subfig:point-forecasting-caltech}
	\end{subfigure}
	\\
	\begin{subfigure}[t]{1.0\linewidth}
		\centering\includegraphics[width=1.0\linewidth]{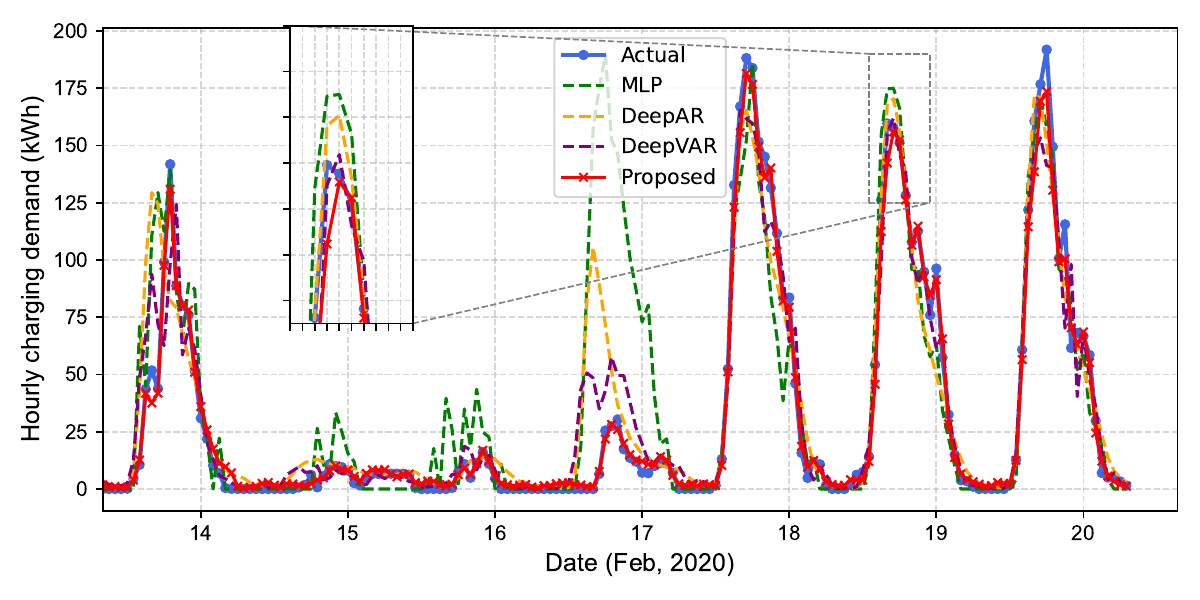}
		\caption{Total Demand.}
		\label{subfig:point-forecasting-total}
	\end{subfigure}
	\caption{Comparison of point forecasting results.}
	\label{fig:point-forecasting}
\end{figure}

\subsection{Point Forecasting Results}
\label{subsec:point-forecasting}

In this subsection, we evaluate the performance of the proposed forecasting model based on LSTM and PICNN. The reconciling model based on DCL is not considered.

Table~\ref{table:point-forecasing} presents the point forecasting metrics in the test set for different forecasting targets and methods. The bold values represents the best results among different methods. It should be noted that the proposed method and MLP both yield multi-horizon forecasts in one single prediction, whereas DeepAR and DeepVAR obtain multi-horizon forecasts through autoregressive iteration. It is shown that the proposed method outperforms the benchmark methods in all point forecasting metrics. For the total EVCS demand, the prediction error of the proposed method is about 30\% of that of DeepVAR.

Fig.~\ref{fig:point-forecasting} shows a one-week comparison of the point forecasting results for the EVCS of \texttt{Caltech}, starting from the first day in the test set (Feb 14, 2022). The forecasted curves in the figure are concatenated by seven consecutive 24-hour forecast results, each spaced 24 hours apart, resulting in continuous curves.

\begin{table*}[!t]
	\renewcommand{\arraystretch}{1.3}
	\caption{Probabilistic Forecasting Evaluation}
	\label{table:probabilistic-forecasting}
	\centering
	\begin{tabu}{ccccccccccccc}
		\toprule
		EVCS & Method & QL(0.1) & QL(0.2) & QL(0.3) & QL(0.4) & QL(0.5) & QL(0.6) & QL(0.7) & QL(0.8) & QL(0.9) & WS(0.6) & WS(0.8)  \\ 
		\midrule
		\multirow{4}{*}{Caltech} &  MLP &   0.9822 &  1.5813 &  2.0293 &  2.3225 &  2.4976 & 2.5058 &  2.3765 &   2.0621 &   1.4736 & 18.2167 & 24.5578 \\
		&       DeepAR &           0.5760 &           0.9783 &           1.3250 &           1.5834 &           1.7478 &           1.7961 &           1.7317 &           1.5203 &           1.0977 &           12.4928 &           16.7369 \\
		&   DeepVAR &           0.6079 &           1.0640 &           1.4076 &           1.6560 &           1.8069 &           1.8407 &           1.7358 &           1.4986 &           1.0793 &           12.8133 &           16.8720 \\
		& Proposed &          \textbf{ 0.1538} &          \textbf{ 0.2480} &           \textbf{0.3283} &           \textbf{0.3953} &           \textbf{0.4500} &           \textbf{0.5510} &           \textbf{1.4552} &           \textbf{1.2154} &           \textbf{0.7460} &            \textbf{7.3171} &            \textbf{8.9976} \\
		\hline
		\multirow{4}{*}{JPL} &          MLP &           2.5150 &           3.4899 &           4.0683 &           4.3627 &           4.4774 &           4.3267 &           3.9518 &           3.3033 &           2.2180 &           33.9659 &           47.3298 \\
		&       DeepAR &           2.1261 &           3.5473 &           4.3583 &           4.7693 &           4.8017 &           4.5420 &           4.0353 &           3.3738 &           2.3449 &           34.6053 &           44.7093 \\
		&   DeepVAR &           1.2319 &           1.9820 &           2.4595 &           2.7196 &           2.8225 &           2.7835 &           2.5820 &           2.1895 &           1.4830 &           20.8572 &           27.1497 \\
		& Proposed &           \textbf{0.4011} &           \textbf{0.6402} &           \textbf{0.8325} &           \textbf{0.9984} &           \textbf{1.1607} &           \textbf{2.7259} &           \textbf{2.5764} &           \textbf{2.1274} &           \textbf{1.3078} &           \textbf{13.8377} &           \textbf{17.0890}  \\
		\hline
		\multirow{4}{*}{Office001} &          MLP &           0.1868 &           0.3818 &           0.5221 &           0.6286 &           0.6841 &           0.6931 &           0.6568 &           0.5701 &           0.3987 &            4.7596 &            5.8552 \\
		&       DeepAR &           0.1799 &           0.3601 &           0.5437 &           0.6485 &           0.7283 &           0.7357 &           0.6895 &           0.5838 &           0.3851 &            4.7198 &            5.6503 \\
		&   DeepVAR &           0.1778 &           0.2761 &           0.3470 &           0.3970 &           0.4265 &           0.4344 &           0.4163 &           0.3644 &           0.2682 &            3.2026 &            4.4597 \\
		& Proposed &          \textbf{ 0.0395} &           \textbf{0.0551} &           \textbf{0.0638} &           \textbf{0.0652} &           \textbf{0.0617} &           \textbf{0.2045} &           \textbf{0.1819} &           \textbf{0.1410} &           \textbf{0.0839} &            \textbf{0.9807} &            \textbf{1.2334} \\
		\hline
		\multirow{4}{*}{Total} &          MLP &           3.6932 &           5.2064 &           6.0464 &           6.5854 &           6.7455 &           6.5437 &           5.9667 &           5.0245 &           3.2964 &           51.1414 &           69.8885 \\
		&       DeepAR &           1.7156 &           2.9709 &           3.8721 &           4.4649 &           4.7767 &           4.7893 &           4.4456 &           3.7235 &           2.4810 &           33.4717 &           41.9662 \\
		&   DeepVAR &           1.6283 &           2.5765 &           3.1543 &           3.5019 &           3.6769 &           3.6923 &           \textbf{3.4996} &           \textbf{3.0019} &           2.0870 &           27.8919 &           37.1521 \\
		& Proposed &           \textbf{0.6751} &           \textbf{0.9715} &           \textbf{1.1591} &           \textbf{1.2365} &           \textbf{1.2654} &           \textbf{1.5652} &           3.6482 &           3.0296 &           \textbf{1.9107} &           \textbf{20.0056} &           \textbf{25.8580} \\
		\bottomrule
	\end{tabu}
\end{table*}

\subsection{Probabilistic Forecasting Results}
\label{subsec:probabilistic-forecasting}

Table~\ref{table:probabilistic-forecasting} presents the probabilistic performance evaluation for the forecasting models. Overall, the proposed method outperforms the other methods for all EVCSs and evaluation metrics, except for QL(0.7) and QL(0.8) on the total demand where DeepVAR performs slightly better. 
For most of the methods, the probabilistic forecasting performance are asymmetric w.r.t. to $ \alpha $, i.e., the quantile functions are difficult to predict at large  $\alpha $ values, due to the skewed nature of the charging demand distribution.

\subsection{Reconciling Model Evaluation}

\begin{figure}[t]
	\centering
	\begin{subfigure}[t]{0.48\linewidth}
		\centering\includegraphics[width=1.0\linewidth]{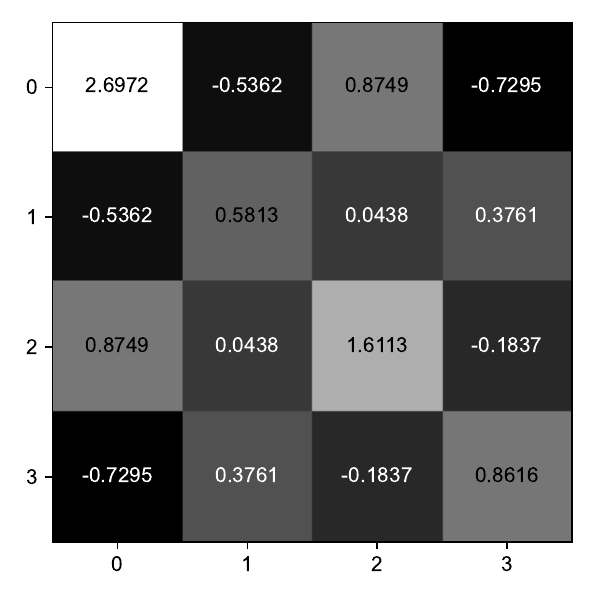}
		\caption{Learned by DCL.}
		\label{subfig:P_opt}
	\end{subfigure}
	~
	\begin{subfigure}[t]{0.48\linewidth}
		\centering\includegraphics[width=1.0\linewidth]{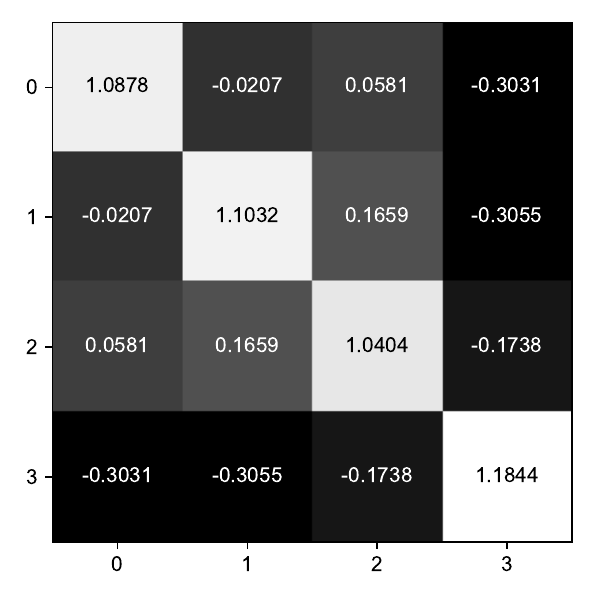}
		\caption{Historical error-based.}
		\label{subfig:P_coef}
	\end{subfigure}
	\caption{Weight adjustment matrices $ \boldsymbol{Q} $.}
	\label{fig:weight-matrix}
\end{figure}

\begin{table}[t]
	\renewcommand{\arraystretch}{1.3}
	\caption{Energy scores and ANOVA test results for different reconciling methods}
	\label{table:energy-score}
	\centering
	\begin{tabular}{ccccc}
		\toprule
		& $ \boldsymbol{Q}_{\text{DCL} } $ & $ \boldsymbol{Q}_{\text{coef} } $ & $ \boldsymbol{Q}_{\text{id} } $ & Original  \\ \midrule
		$ L_{\text{ES}}  $ & \textbf{14.2982}  &  15.7225  &  15.3431  & 17.2015 \\
		\bottomrule
	\end{tabular}
	\\ \vspace{1em}
	\begin{threeparttable}
		\begin{tabular}{cccc}
			\toprule
			\thead{F statistic\\(P-value)} & $ \boldsymbol{Q}_{\text{DCL} } $ & $ \boldsymbol{Q}_{\text{coef} } $ & $ \boldsymbol{Q}_{\text{id}} $ \\
			\midrule
			$ \boldsymbol{Q}_{\text{coef} } $ & \thead{11.4901\\(7.19e-4)***} & &  \\
			$ \boldsymbol{Q}_{\text{id}} $ & \thead{6.1978\\(1.29e-2)*} & \thead{0.7999\\(0.37)} & \\
			Original & \thead{44.1556\\(4.34e-11)***} & \thead{11.2367\\(8.23e-4)***} & \thead{17.7763\\(2.64e-5)***}  \\
			\bottomrule
		\end{tabular}
		\begin{tablenotes}[flushleft]
			\item[\dag] Overall F statistic: 15.58, P-value: 4.74e-10 ***.
		\end{tablenotes}
	\end{threeparttable}
\end{table}

The forecasted results in Subsections~\ref{subsec:point-forecasting} and~\ref{subsec:probabilistic-forecasting} are obtained directly from the base forecasting models, and they do not necessarily satisfy the \textit{hierarchical constraint}. 1,000 scenarios are sampled from the proposed method based on LSTM and PICNN (ReLU+Gaussian softplus), and these base samples are reconciled using DCL. In the proposed framework, the weight matrix $ \boldsymbol{Q} $ in~(\ref{equ:hierarchical-reconciling}) are learned from the energy score in the original validation set ($ \boldsymbol{Q}_{\text{DCL}} $), as described in Subsection~\ref{subsec:DCL-for-reconciling}. For comparison, we also tested the reconciled results with $ \boldsymbol{Q} $ estimated in two traditional ways: i) $ \boldsymbol{Q} $ is the inverse of the correlation matrix for the base forecasting errors in the validation set ($ \boldsymbol{Q}_{\text{coef}} $); ii)  $ \boldsymbol{Q} $ is the identity matrix ($ \boldsymbol{Q}_{\text{id}} $).
Fig.~\ref{fig:weight-matrix} shows the $ \boldsymbol{Q} $ matrices for $ \boldsymbol{Q}_{\text{DCL}} $ and $ \boldsymbol{Q}_{\text{coef}} $. 

Table~\ref{table:energy-score} shows the energy scores for different reconciling methods in the test set. The energy score from the reconciled scenarios based on $ \boldsymbol{Q}_{\text{DCL}} $ is the lowest among all the methods, and is reduced to about 83\% of that from the original scenarios. For each time interval in the test set, four series of energy scores are calculated, followed by statistical analysis using the ANOVA test to determine the significance of the reduction in these scores. The overall and pairwise F statistics and P-values are also presented in the Table. The proposed method significantly outperforms other methods in terms of energy score. It is also shown that the reconciled results from $ \boldsymbol{Q}_{\text{coef}} $ and $ \boldsymbol{Q}_{\text{id}} $ have similar performance. 

\begin{table}[t]
	\renewcommand{\arraystretch}{1.3}
	\caption{MAE before and after reconciliation}
	\label{table:reconciled-MAE}
	\centering
	\begin{tabular}{ccccc}
		\toprule
		Reconciliation & Caltech &  JPL &   Office001 &   Total \\
		\midrule
		$ \boldsymbol{Q}_{\text{DCL} } $ & 0.8012 & \textbf{1.7772} &      \textbf{0.1657} &  \textbf{1.9614} \\
		$ \boldsymbol{Q}_{\text{coef} } $ & 1.1049 & 1.8786 &      0.6454 &  2.0308 \\
		$ \boldsymbol{Q}_{\text{id} } $ & 1.0159 & 1.9148 &      0.5804 &  1.9799 \\
		Original & \textbf{0.7981} & 2.3712 &      0.1825 &  2.1546 \\
		\bottomrule
	\end{tabular}
\end{table}

\begin{figure}[t]
	\centering
	\includegraphics[width=0.8\linewidth]{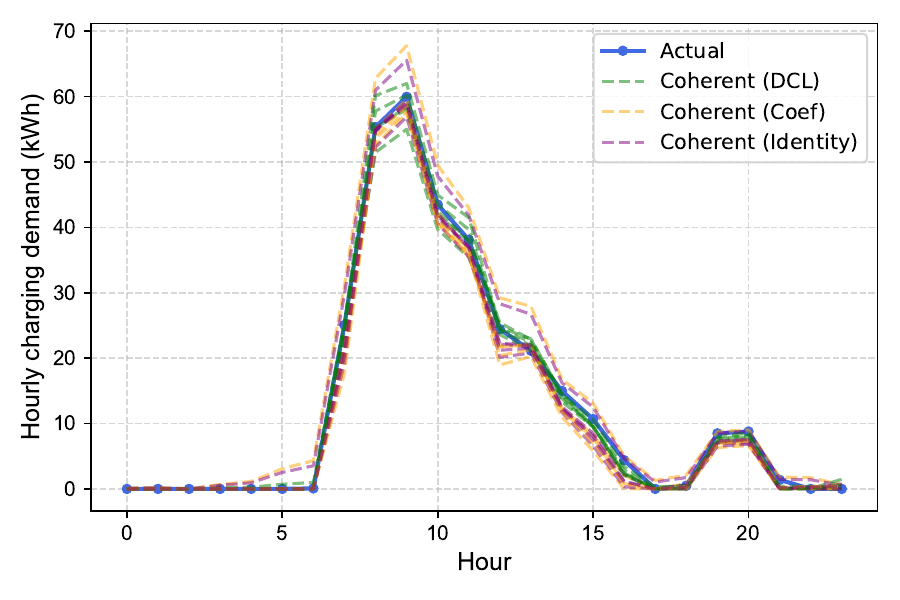}
	\caption{Comparison of stochastic scenarios for Feb 18, 2020.}
	\label{fig:coherent-scenario}
\end{figure}

Table~\ref{table:reconciled-MAE} further presents the point forecasting error before and after reconciliation. The proposed reconciling method based on DCL can reduce the the overall MAE for the four sites by about 14.6\%. 
The proposed method dominates all the comparisons in MAE for all the targets except \texttt{Caltech} (for which the magnitude of the difference between the proposed method and the original is almost negligible), indicating that the quality of the predicted scenarios is improved through reconciliation while maintaining high forecasting accuracy. 

Fig.~\ref{fig:coherent-scenario} plots 5 samples for each reconciling method. The scenarios obtained from $ \boldsymbol{Q}_{\text{DCL}} $ are more compact and close to the actual demand, in comparison to the other methods.


\subsection{Sensitivity Analysis}

Practically, a longer forecasting horizon might be needed for EVCS operators' decision. In addition to the aforementioned 24-hour ahead forecasting results, 48-, 72-, and 96-hour ahead forecasting are conducted as sensitivity analysis. To keep the conciseness of this paper, we only present results of a portion of the evaluation metrics for \texttt{Total}. The results of MLP are also omitted due to its underperformance compared to other benchmark methods. 

Table~\ref{table:metrics-sensitivity} shows some point and probabilistic evaluation metrics for different horizon lengths. It is usually more difficult to forecast a larger horizon. The proposed method still outperforms the comparisons to a significant extent. 
Table~\ref{table:energy-score-sensitivity} shows the energy scores for sensitivity analysis. It is noteworthy that the reconciled results from $ \boldsymbol{Q}_{\text{coef}} $ and $ \boldsymbol{Q}_{\text{id}} $ might be worse than the original scenarios in 72- and 96-hour ahead forecasting. The scenarios obtained from $ \boldsymbol{Q}_{\text{DCL}} $ are still better, although the improvement becomes less as the horizon length increases.

\begin{table}[h]
	\renewcommand{\arraystretch}{1.3}
	\caption{Evaluation metrics for different lengths of the forecasting horizons}
	\label{table:metrics-sensitivity}
	\centering
	\begin{tabular}{cccccc}
		\toprule
		Method &  \thead{Horizon\\Length} & MAE & QL(0.2) &  QL(0.8) &  WS(0.6)  \\
		\midrule
		\multirow{3}{*}{DeepAR} &   48 &   9.1122 &        2.9300 &           3.4255 &           31.7778  \\
		 &   72 &     9.9185 &         3.1920 &        3.7305 &           34.6124  \\
		 &   96 &     10.3336 &        3.2585 &         3.8273 &           35.4292  \\ \hline
		\multirow{3}{*}{DeepVAR} &   48 &    8.6640 &       3.1911 &           3.6232 &           34.0717  \\
		 &   72 &      7.7221 &       2.7857 &         3.2860 &           30.3585  \\
		 &   96 &        8.9596 &       3.3804 &       3.9103 &           36.4534  \\ \hline
		\multirow{3}{*}{Proposed} &   48 &    5.6690 &       2.5432 &           2.9229 &           27.3301  \\
		 &   72 &      5.4981 &        2.2896 &        2.9005  &           24.5254  \\
		 &   96 &      6.7632 &      2.9280 &        2.8279 &           28.5281 \\
		\bottomrule
	\end{tabular}
\end{table}

\begin{table}[h]
	\renewcommand{\arraystretch}{1.3}
	\caption{Energy scores for different reconciling methods under different lengths of the forecasting horizon}
	\label{table:energy-score-sensitivity}
	\centering
	\begin{tabular}{ccccc}
		\toprule
		\thead{Horizon\\Length} & $ \boldsymbol{Q}_{\text{DCL} } $ & $ \boldsymbol{Q}_{\text{coef} } $ & $ \boldsymbol{Q}_{\text{id} } $ & Original  \\ \midrule
		48  & \textbf{36.5978}  &  45.7718  &  44.3981  & 49.1750 \\
		72  & \textbf{56.6180}  &  63.9498  &  61.1891  & 63.4553 \\
		96  & \textbf{91.2865}  &  99.4984  &  95.0290  & 94.9913 \\
		\bottomrule
	\end{tabular}
\end{table}

\subsection{Time Consumption of DCL}

We also recorded the time consumption of DCL w.r.t. the length of the forecasting horizon. For each epoch of the training set, the average time consumption for the forward pass and gradient calculation of DCL in 24-hour ahead forecasting is 61.97 seconds and 27.97 seconds, respectively. The average time consumption for 48-hour, 72-hour, and 96-hour ahead forecasting is 74.29/29.55, 102.56/31.55, 115.65/33.77 seconds, respectively. The results indicate that DCL is still efficient even the length of the forecasting horizon is increased to 96.

\section{Conclusion}
\label{sec:conclusion}
This paper proposes a novel deep learning-based framework for short-term hierarchical probabilistic forecasting for EVCS demand. LSTM is used as the forecasting engine, and convex learning layers of PICNN and DCL are used for capturing the probabilistic distribution and reconciling the stochastic scenarios, respectively. 
The conditional quantile function of the target variable is learned by PICNN, and the weight matrix for hierarchical adjustment is learned by DCL.
Real world EV charging data from the open source ACN dataset are used to illustrate the procedures of the proposed framework and demonstrate the performance in terms of forecasting accuracy and scenario quality. 

Future work includes testing the method in more complicated hierarchical structures.

\appendix  

\label{sec:appendix}
\setcounter{figure}{0}
\setcounter{table}{0}
\renewcommand\thefigure{A\arabic{figure}}
\renewcommand\thetable{A\arabic{table}}

A brief discussion on the selection of activation functions for PICNN is given here. 
The number of hidden layers and combination of activation functions have significant impact on the fitting and generalization of PICNN. 
Typically, the performance on the validation set is used as the primary criterion for the selection of the structure, but intuitive results on the shape of the fitted curve of the conditional CDF can also help. For simplicity, we mainly focus on two types, ReLU (r) and Gaussian softplus (g). 

\begin{figure}[h]
	\centering
	\begin{subfigure}[t]{1.0\linewidth}
		\centering\includegraphics[width=0.7\linewidth]{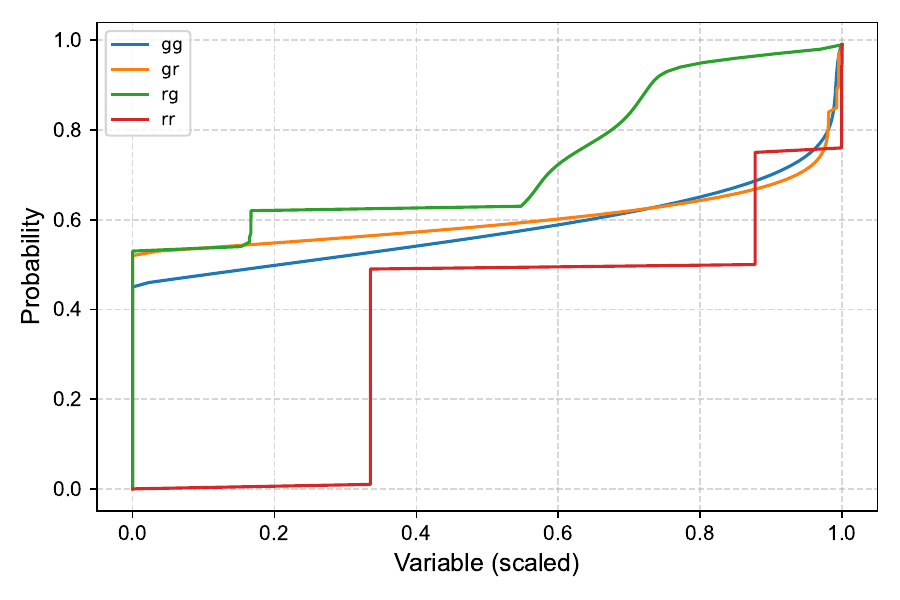}
		\caption{Two hidden layers.}
		\label{subfig:CDF-2}
	\end{subfigure}
	\\
	\begin{subfigure}[t]{1.0\linewidth}
		\centering\includegraphics[width=0.7\linewidth]{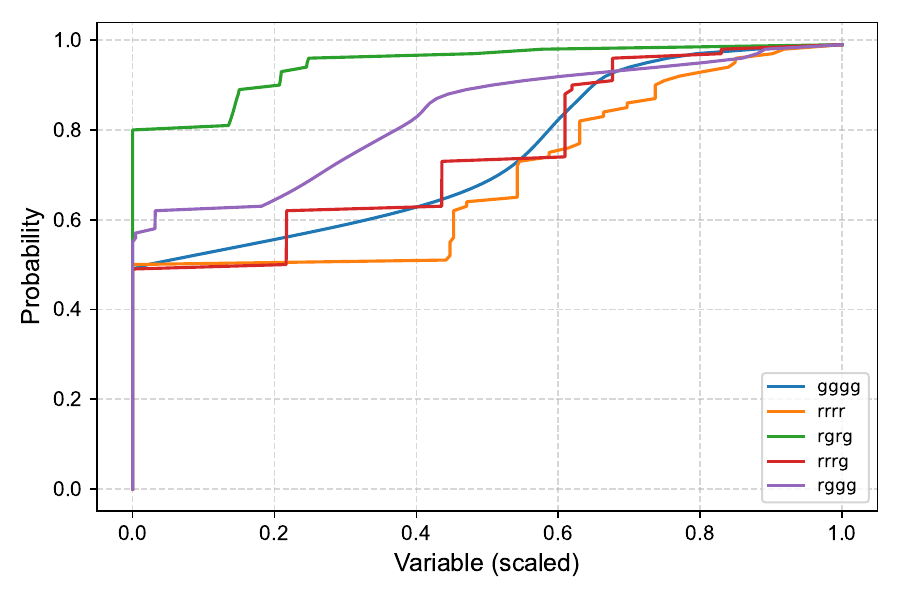}
		\caption{Four hidden layers.}
		\label{subfig:CDF-4}
	\end{subfigure}
	\caption{Scaled conditional CDFs for different combination of activation functions for PICNN.}
	\label{fig:CDF}
\end{figure}

Fig.~\ref{fig:CDF} shows the scaled conditional CDFs for different combination of activation functions for the prediction of the Caltech EVCS demand at one single time period. The variable is scaled between $ [0,1.0] $. We only plot the cases of two and four hidden layers due to space limitation. It can be seen that the ReLU layer increases the piecewise fitting ability, while the Gaussian softplus layer increases the smoothness of the predicted conditional CDF. 
Table~\ref{table:activation-combination} presents the average MAE in the validation set for different structures of PICNN. The bold values represent the best values for each specific number of hidden layers.
It is shown that using hybrid ReLU and Gaussian softplus layers has the best performance regardless of the number of hidden layers. Increasing layers does not significantly reduce the validation loss, and the best structure is ReLU+Gaussian softplus, which is further used in the case study.

\begin{table}[h]
	\renewcommand{\arraystretch}{1.3}
	\caption{Average validation MAE for different combination of activation functions}
	\label{table:activation-combination}
	\centering
	\begin{tabu}{cc|cc}
		\toprule
		Activation function &      MAE & Activation function &    MAE \\
		\midrule
		gg & 1.8633 &                  ggrg & 2.7462 \\
		gr & 1.8110 &                  ggrr & 1.7077 \\
		rg & \textbf{1.5444} &                  grgg & 1.8349 \\
		rr & 1.7505 &                  grgr & 1.9137 \\ \cline{1-2}
		ggg & 2.4505 &                  grrg & 2.8953 \\ 
		ggr & 2.9051 &                  grrr & 1.9123 \\
		grg & 1.9013 &                  rggg & 2.0963 \\
		grr & 2.3299 &                  rggr & 2.5571 \\
		rgg & 2.4048 &                  rgrg & 2.3689 \\
		rgr & 1.9127 &                  rgrr & 2.2289 \\
		rrg & \textbf{1.6798} &                  rrgg & 2.3937 \\
		rrr & 1.7352 &                  rrgr & \textbf{1.5456} \\ \cline{1-2}
		gggg & 2.6010 &                  rrrg & 2.0953 \\
		gggr & 3.2011 &                  rrrr & 1.8318 \\
		\bottomrule
	\end{tabu}
\end{table}


%



\bibliographystyle{IEEEtran}
\bibliography{ref}

\begin{IEEEbiography}[{\includegraphics[width=1in,height=1.25in,clip,keepaspectratio]{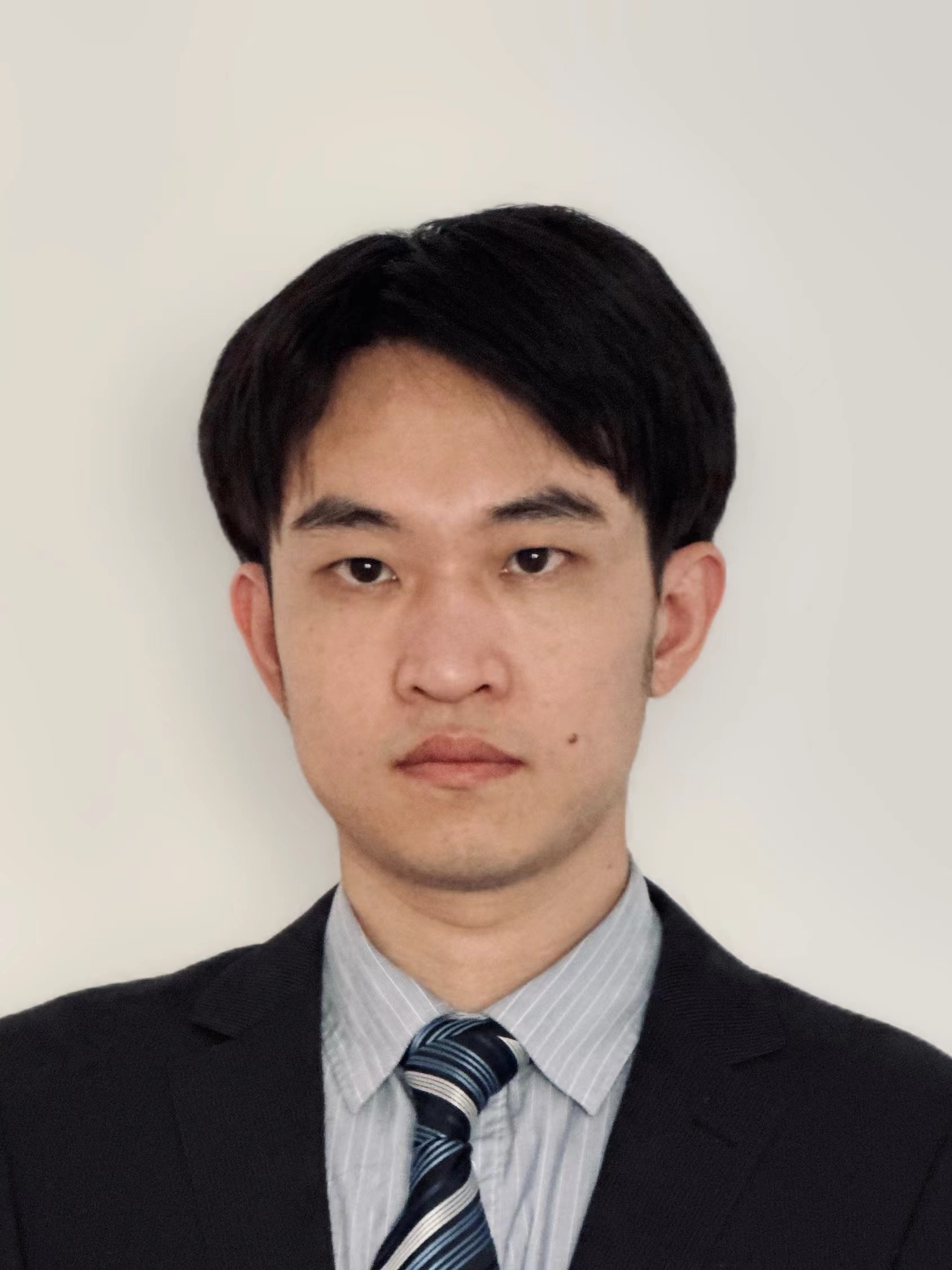}}]{Kedi Zheng} (Member, IEEE) 
	received the B.S. and Ph.D. degrees in electrical engineering from Tsinghua University, Beijing, China, in 2017 and 2022, respectively.
	
	He is currently a Post-Doctoral Researcher with Tsinghua University. He was also a Visiting Research Associate with The University of Hong Kong in 2023. His research interests include data analytics in power systems and electricity markets.
\end{IEEEbiography}

\begin{IEEEbiography}[{\includegraphics[width=1in,height=1.25in,clip,keepaspectratio]{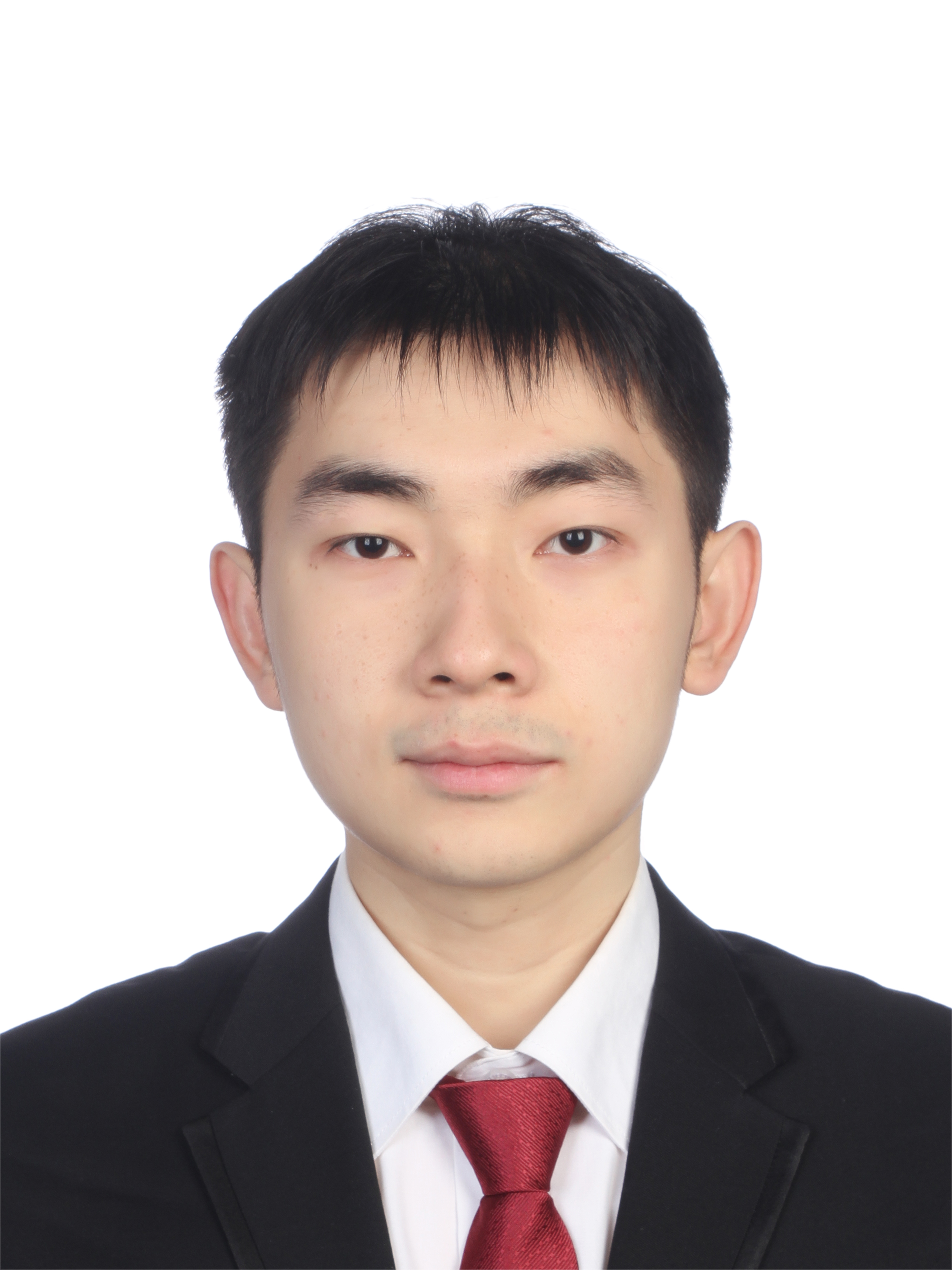}}]{Hanwei Xu} 
	received the B.S. and M.S. degrees in electrical engineering from Tsinghua University, Beijing, China, in 2017 and 2020, respectively. 
	
	He is currently a Researcher with Beijing Deepseek Artificial Intelligence Fundamental Technology Research Co., Ltd. He was a Visiting Graduate Student at the University of Nottingham in 2019.
	His research interests include neural networks, deep learning and natural language processing.
\end{IEEEbiography}

\begin{IEEEbiography}[{\includegraphics[width=1in,height=1.264in,clip,keepaspectratio]{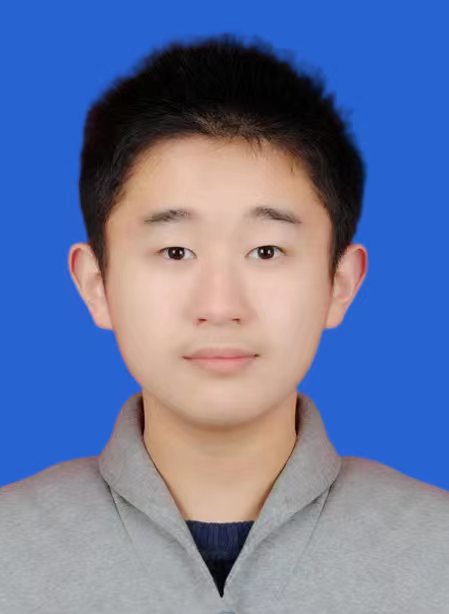}}]{Zeyang Long} is currently working toward a B.S. degree in electrical and electronic engineering at Huazhong University of Science and Technology. His research interests include deep learning, power market, and energy forecasting.
\end{IEEEbiography}

\begin{IEEEbiography}[{\includegraphics[width=1in,height=1.264in,clip,keepaspectratio]{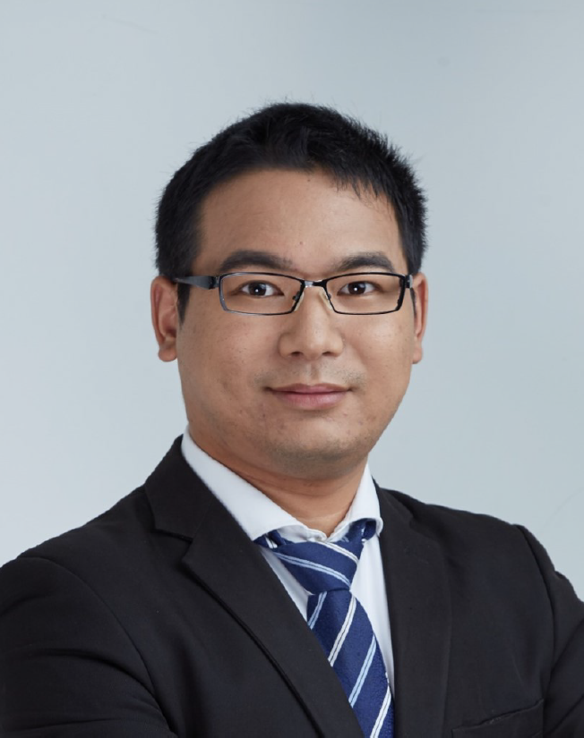}}]{Yi Wang} (Member, IEEE) received the B.S. degree from Huazhong University of Science and Technology in June 2014, and the Ph.D. degree from Tsinghua University in January 2019. He was a visiting student with the University of Washington from March 2017 to April 2018. He served as a Postdoctoral Researcher in the Power Systems Laboratory, ETH Z\"{u}rich from February 2019 to August 2021. 
	
He is currently an Assistant Professor with the Department of Electrical and Electronic Engineering, The University of Hong Kong. His research interests include data analytics in smart grids, energy forecasting, multi-energy systems, Internet-of-things, cyber-physical-social energy systems.
\end{IEEEbiography}

\begin{IEEEbiography}[{\includegraphics[width=1in,height=1.25in,clip,keepaspectratio]{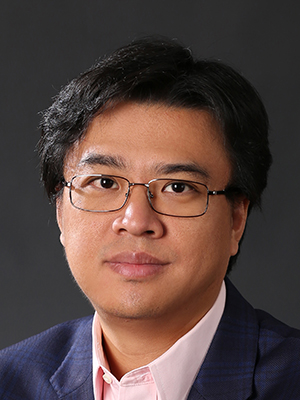}}]{Qixin Chen} (Senior Member, IEEE)
	received a Ph.D. degree from the Department of Electrical Engineering at Tsinghua University, Beijing, China, in 2010.
	
	He is currently a Professor at Tsinghua University. His research interests include electricity markets, power system economics and optimization, low-carbon electricity and power generation expansion planning.
\end{IEEEbiography}

\vfill
\end{document}